\documentstyle[12pt,epsfig,a4]{article}
\begin{document}

\date{June 1995}
\title
{\bf Study of the Strongly Interacting Higgs Physics Model
with Heavy Fermions }
\author
{L.V.~Dung${}^{1}$, G.~Jikia${}^{2}$, C.~Roiesnel and T.N.~Truong \\
{\normalsize{\it Centre de Physique Th\'eorique,}}\\
{\normalsize{\it Centre National de la Recherche Scientifique, UPR A0014,}}\\
{\normalsize{\it Ecole Polytechnique, 91128 Palaiseau Cedex, France}} }
\maketitle
\begin{abstract}
We study the effect of a possible fourth heavy generation of fermions
on the Higgs sector of the standard model. We show, from the violation
of elastic unitarity, that the scale of strong interactions is well
below 1 TeV even with a Higgs mass as low as 500 GeV provided the
fourth generation fermion mass is equal or larger than the Higgs mass.
The diagonal Pade approximant method is then used to unitarize
the partial wave amplitudes. It is found that, for the fourth generation
fermion masses which are comparable to or larger than the Higgs mass,
the Landau ghosts in the I=0 and  I=2 channels of the reconstructed
amplitudes move too close to the physical region to be accepted.
\end{abstract}
\footnotetext[1]{Permanent address: {\it Theoretical Physics Institute,
BP429 BoHo, Hanoi 10000, Vietnam.}}
\footnotetext[2]{Permanent address: {\it Institute for High Energy Physics,
142284, Protvino, Moscow Region, Russia.}}
\vskip 5.7cm
\leftline{\normalsize RR359.0695}
\newpage

\section{Introduction}

        The standard model is verified to a great accuracy by recent
experiments at LEP. The symmetry breaking of the electroweak interaction
is however less understood because its effect is insensitive to these
low energy experiments. This is the consequence of Veltman's screening
theorem \cite{Velt}. Next generation experiments at LHC are
designed to probe the Higgs sector. If the
Higgs boson is sufficiently heavy, the study of the longitudinal
vector boson scattering can provide valuable informations on the
spontaneous symmetry breaking mechanism of the standard model. The
theoretical study of this process is a challenge to theoreticians
because the standard perturbative field theoretical methods have to be
modified due to the strong interaction of the heavy Higgs sector. This
problem is well-known and has recently been the focus of extensive
studies in the litterature \cite{XXX} based on the equivalence theorem
\cite{EQT} with the linear sigma model \cite{LSM}. 

        An equally interesting problem, but less well-known, is to
consider  the effect
of a possible heavy fourth generation of fermions \cite{DV}. The
existence of heavy weak $SU(2)$ multiplets, sufficiently
degenerate to satisfy the $\rho$ parameter constraint, is still
consistent with experimental data. Heavy fermions
interact strongly because of large Yukawa couplings and along with the
Higgs boson represent strongly interacting quanta of the symmetry breaking
sector. This hypothesis will
introduce another scale in the problem and physical amplitudes, even
 at low energy, does not obey the screening theorem, i.e. the Higgs
 and fermion mass dependence of physical quantities is no longer logarithmic
 but rather power like. This result is related to Hill's
conjecture \cite{Hill} about the non-decoupling of heavy quanta when
adding extra strong interactions in the Higgs sector. In a previous
preliminary study of the consequences of the existence of a fourth generation
of fermions \cite{Truong}, it was found that
the P-wave elastic scattering amplitude of the longitudinal
$W$ bosons depends sensitively on the ratio of the Higgs and fermion masses.
The purpose of this letter is to make a more general and precise
study of this  question.

        Our strategy is to calculate numerically the one-loop $W_{L}W_{L}$
partial-wave amplitudes in all  isospin channels. Using
the Argand diagram analysis we show, from the violation of elastic unitarity,
that the strong interaction physics sets in even with a Higgs mass as
low as 500 GeV provided the 4th generation fermion mass is equal or larger than
the Higgs mass. We then make the hypothesis that the elastic
unitarity  condition at low energy, where  inelasticity is negligible,
is an important condition which must be satisfied by strong
interactions. This leads us to unitarize the
pertubative amplitudes by the Pad\'e approximant method \cite{TNT}. 
We also require that the unitarisation scheme should be
consistent with analyticity or causality or, in other words, the
presence of  the Landau ghosts due to the unitarisation scheme should
not be  accepted if they are too close to the low energy region.
We study this Landau ghost problem using the
low energy expansion of the one loop amplitude and find that the
solutions with the fourth generation
fermion masses which are comparable or larger than the Higgs mass, the  Landau
ghosts in the $I=0$ and $I=2$ channels  move too close to the physical
region to be accepted.

        Our main conclusion is that if the ratio of the Higgs to
 the fermion mass is less or equal to unity, our requirement cannot be
 satisfied with any known non perturbative methods. Large values of the ratio
 of Higgs to fermion mass are entirely acceptable. Although we work
with a heavy  Higgs assumption so that the equivalence theorem can be
applied, which simplifies considerably our calculation, we suspect that
this difficulty is also present in a scenario with a relatively light
Higgs of a few hundred GeV but with heavier fermions.

\section{Argand plots for $W_{L}W_{L}$ scattering}

        The real parts of the one-loop corrections to longitudinal vector boson
scattering due to a heavy fourth generation of fermions have been calculated
by Dawson and Valencia \cite{DV} using the Goldstone boson equivalence
theorem \cite{EQT} and the
On-Mass-Shell renormalization scheme. Their calculation can be shown
to be in agreement with a much earlier one by Jhung and Willey
\cite{Willey}. For a numerical evaluation by the 
FF package \cite{FF}, we find it convenient to express the complex
$W_{L}W_{L}$ scattering amplitudes in terms of the  4-, 3- and 2-point
scalar loop integrals. When the fermions are
degenerate, we need only to calculate the one-loop amplitude $A(s,t,u)$ for the
process $W_{L}^{+}W_{L}^{-}\rightarrow Z_{L}Z_{L}$. We get
$$A(s,t,u) = A_{b} + A_{f}$$
where $A_{b}$ and $A_{f}$, respectively the bosonic and fermionic
contribution, read
\begin{eqnarray}
\lefteqn{
16 \pi^2 v^4\, A_b=
m_H^8\biggl\{ {\it D_{wHwH}}(t,s)+{\it D_{wHwH}}(u,s)\biggr\}
+2 m_H^6\biggl\{ {\it C_{wHw}}(t)+2\, {\it C_{wHw}}(u)\biggr\}
}\nonumber \\
&&
+\frac{2\,m_H^{6}}{s-m_H^{2}}\Biggl\{
s\,{\it C_{wHw}}(s)+\left (s+2\,m_H^{2}\right ){\it C_{HwH}}(s)
+2\,{\it B_{wH}}(0)+s\, {\it  B'_{wH}}(0)
\Biggr\}
\nonumber \\
&&
+\frac{m_H^4}{2\, (s-m_H^2 )^2}
\Biggl\{s \left(7s-4m_H^{2}\right ){\it B_{ww}}(s)
- 3 s^2\,{\it \Re e B_{ww}}(m_H^{2})
-9{s}^2 {\it B_{HH}}(m_H^{2})
\nonumber \\
&&
+\left (s+2\,m_H^{2}\right )^{2}{\it B_{HH}}(s)
\Biggr\}
+m_H^4\biggl\{{\it B_{ww}}(t)+{\it B_{ww}}(u)\biggr\}
-{\frac {m_H^4s}{2\,(s-\,m_H^{2})}}
\label{boson}\\
\lefteqn{
\frac{16\pi^2 v^4}{N_c}\, A_f =
4\,M_f^4
\biggl \{st\,{\it D_{ffff}}(s,t)+su\,{\it D_{ffff}}(s,u)-tu\,{\it D_{ffff}}(t,u)
\biggr\}
}\nonumber \\
&&
-{\frac {16\,M_f^{4}{s}^{2}}{
\left (s-m_H^{2}\right )}}{\it C_{fff}}(s)
-\frac{4\, M_f^2}{\left (s-m_H^2\right )^2}
\Biggl\{
\left (m_H^{2}-4\,M_f^{2}\right )
{s}^{2}\, {\it \Re e B_{ff}}(m_H^{2})
\nonumber \\
&&
+ 4\,M_f^{2}{s}^{2}{\it B_{ff}}(s)
-m_H^4\, s \left ({\it B_{ff}}(s)-\,{\it B_{ff}}(0)\right )
-m_H^2\,s^2\,{\it B_{ff}}(0) \Biggr\}
\label{fermion}
\end{eqnarray}
and the scalar loop integrals, denoted as $D$, $C$ and $B$, are
defined by the standard expressions
\begin{eqnarray}
&&{\it D_{wHwH}}(t,s) =
\nonumber\\
&& \frac{1}{i\pi^2}\int 
\frac{dQ}{Q^2((Q+p_3)^2-m_H^2)(Q+p_2+p_3)^2((Q+p_1+p_2+p_3)^2-m_H^2)},
\nonumber\\
&&
{\it C_{wHw}}(s) = \frac{1}{i\pi^2}\int 
\frac{dQ}{Q^2((Q+p_1)^2-m_H^2)(Q+p_1+p_2)^2},
\nonumber\\
&&{\it B_{mM}}(s) = -\int^1_0 dx\, \log(x m^2+(1-x) M^2 -x(1-x)s),
\nonumber\\
&&{\it B'_{mM}}(s)= \frac{d{\it B_{mM}}(s)}{ds}, \quad \mbox{etc.},
\nonumber
\end{eqnarray}
with $s=(p_1+p_2)^2$, $t=(p_2+p_3)^2$, $u=-s-t$.

        The scattering amplitudes $A_{I}(s,t)$ of the  isospin $I=0,1,2$
eigenstates are
\begin{eqnarray}
&& A_{0}(s,t)=3\thinspace A(s,t,u)+A(t,s,u)+A(u,t,s) \nonumber\\
&& A_{1}(s,t)=A(t,s,u)-A(u,t,s)\nonumber\\
&& A_{2}(s,t)=A(t,s,u)+A(u,t,s)\nonumber
\end{eqnarray}
and the partial-wave amplitudes $a_{IJ}(s)$ are defined by
$$a_{IJ}(s)=\frac{1}{64\pi}\int_{-1}^{+1}d(\cos\theta)\thinspace
P_{J}(\cos\theta)\thinspace A_{I}(s,\cos\theta)\,,\quad t=-\frac{s}{2}(1-\cos\theta)$$
        Given these exact one-loop partial-wave amplitudes, one may
try to estimate, as a function of the Higgs mass $m_{H}$ and heavy fermion mass
$m_{f}$, the scale $\sqrt{s}$ at which the standard model becomes
strongly interacting in the Higgs sector. Violation of unitarity can be
a useful guide to limit the region of validity of the one-loop
calculation. For that purpose, we use the Argand plots of
partial-wave amplitudes to find out the degree of violation of elastic
unitarity which can be ascribed to the omission of higher  order terms
in the pertubation series because this expansion only
obeys unitarity order by order.
        
        Let $a_{IJ}^{(0)}$ and $a_{IJ}^{(1)}$ be respectively the
tree-level contribution and the  one-loop corrections to the
partial-wave amplitude $a_{IJ}(s)$. Then one can quantify the size of
unitarity violation \cite{DJ} by the vector $\Delta a_{IJ}$ of minimum length
which can be added to $a_{IJ}=a_{IJ}^{(0)}+a_{IJ}^{(1)}$, to bring the elastic
scattering amplitude to the unitarity circle
$$\mid \Delta a_{IJ}\mid = \left|\frac{1}{2}-\sqrt{\left(a_{IJ}^{(0)}+\Re
a_{IJ}^{(1)}\right)^2+\left(\frac{1}{2}-\Im a_{IJ}^{(1)}\right)^2}\right|$$
There is some arbitrariness in choosing a criterion for satisfactory
convergence of the pertubative expansion. We impose
$$\left| \frac{\Delta a_{IJ}(s)}{a_{IJ}(s)}\right| \leq \frac{1}{2}$$
The energy scales $\Lambda_{c}=\sqrt{s}$ which violate this bound are
gathered in Table~\ref{scales} for a representative set of Higgs and
fermion masses. Argand plots for $I=0$, $J=0$ channel are shown in Fig.~1.

Imposing other criteria such as
${\displaystyle \left|\frac{a_{IJ}^{(1)}}{a_{IJ}^{(0)}}\right| < 1}$, or
${\displaystyle |\Delta a_{IJ}| < \frac{1}{2}}$, would lower the scales
$\Lambda_{c}$. There are no other assumptions than those which
ensure the applicability of the equivalence theorem and the elastic
unitarity condition, namely $M_{W} \ll \sqrt{s} < 2\inf(m_{H},m_{f})$
\footnote{Of course the condition $\sqrt{s}<2m_{H}$ is relaxed
in isospin channels $I\neq 0$.}.
The influence of the top quark with a mass smaller or equal to 200 GeV has negligible effect on the $W_{L}W_{L}$ scattering. The
effect of multiple W production has recently been calculated in the heavy
 Higgs limit \cite{DJ} and was shown to have negligible effect on the
unitarity relation. In the region of energy of interest we can therefore
safely ignore inelastic effects in the unitarity relation.

One can read from Table~\ref{scales} that fermions induce strong
interactions below 1 TeV in the $I=J=0$ channel already for Higgs and
fermion masses between 0.5 GeV and 1 GeV, whereas in the absence of
fermions, interactions become strong below 2 TeV only when the Higgs mass is
greater or equal to 3 TeV \cite{DJ}.

\section{Pad\'e approximant for $W_{L}W_{L}$ scattering}

        In this section we want to study the $W_{L}W_{L}$ scattering problem
by resumming the one-loop calculation in order to make it satisfy
the unitarity relation. This is necessary in order to extend the region
of validity of the perturbation calculation as this method can handle
strong interactions and/or resonance effects. The exact preservation of
probability is  utmost important to handle these problems.

        Let us denote the $l^{th}$ partial wave projection of the perturbative
amplitude by $f$ where the isospin and $l^{th}$ partial wave indices
and the $s$ dependence are
omitted. The corresponding tree and one loop amplitude are denoted
respectively by $f^{(0)}$ and $f^{(1)}$. The perturbative amplitude
$f =f^{(0)} + f^{(1)}$ satisfies the perturbative unitarity relation,
$Imf =\left( f^{(0)} \right)^{2}$,
which is to be compared with the exact elastic unitarity relation,
$Imf =( f )^{2}$, which must be valid in the energy region below the
inelastic threshold. 
The one-loop perturbative series can be resummed by consructing the
diagonal $[1,1]$  Pad\'e approximant:
\begin{eqnarray}
\label{pade}
f^{[1,1]} =\frac{ f^{(0)}}{1 - f^{(1)}/f^{(0)}}
\end{eqnarray}
which satisfies the elastic unitarity relation 
$Im f^{[1,1]} = \mid f^{[1,1]} \mid^{2}$. Hence we can write 
$f^{[1,1]} = e^{i\delta}\sin\delta$, where $\delta$ is the phase shift.

        If we add any real polynomial contribution to the denominator of
Eq~\ref{pade} or any analytic function which is real on the unitarity cut,
the elastic unitarity relation remains valid. Hence the Pad\'e
method can also be used to study the inelastic effect of the $HH$ and $FF$
intermediate states on the elastic scattering provided we limit ourself to the
energy region below the inelastic threshold.

        In Fig.~2 some typical phase shifts are plotted for Higgs boson masses
$m_H=1$ and 1.5~TeV as a
function  of energy. Because of our unitarisation method, for a Higgs
mass  less than 3 TeV, the Higgs mass defined by the position where
the $I=J=0$ phase shift passing through $90^{\circ}$ is the same as
that defined by the vanishing of the real part of the inverse of the
Higgs propagator. For a  Higgs mass larger than 3 TeV a lower energy
resonance is generated with a phase shift passing through $90^{\circ}$ from
below and is genuinely a bonafide resonance. The Higgs mass that we
defined and used in our perturbative calculation corresponds  in
fact to a $I=J=0$ phase shift passing through $90^{\circ}$ from above.

        For $m_{H}^{2}/m_{f}^{2}\leq 1$ the $I=2$ S-wave phase
shift also exhibits  a resonance behavior but with the phase shift
passing through $-90^{\circ}$.  We show below it is associated with
a Landau ghost very close to the real energy axis and cannot be
accepted. It is generated by the unitarisation scheme in which the
scattering length and the effective range, respectively the
$O(p^{2})$ and $O(p^{4})$ terms, are both negative due the
contribution of the fermion loop in this channel.

        The $I=J=1$ phase shift has a resonance behavior ($\rho$). The
position of the resonance is much lower than that given by technicolor
models \cite{Truong}. Furthermore the KSRF relation is not valid. Our
calculation shows that  for small values of Higgs mass, although the
analytical  results of \cite{Truong} are qualitatively correct, a
substantial correction to the width of $\rho$  due to the contribution
of the $t\bar{t}$ channel must be made. This is so because the $\rho$
width due to the $W_{L}W_{L}$ states are so small that even the
$t\bar{t}$ contribution becomes dominant. 

\section{Some problems associated with unitarisation}

        The unitarisation scheme does not always work because it
may introduce unwanted  singularities in the complex energy plane which
give rise to a violation of causality. These Landau type singularities
can be got rid of by multiplying the scattering amplitude by an appropiate
polynomial. This can be obviously done but
at the expense of a violation of unitarity. If they were far from the
physical region of interest, the resulting violation of unitarity would
be small and could be blamed on our approximation scheme and therefore could
be accepted.  Therefore we should examine
carefully the unitarised amplitudes to see whether they have unwanted
singularities which are too close to the physical region of interest.

        However our Pad\'e amplitude is computed numerically for real
energy and is not suitable for analytic continuation to find complex
singularities. A more exact calculation of the Landau ghost resulting
from our Pad\'e amplitude will be undertaken shortly and will be a
subject for a future publication. In the meanwhile we can use the
$s\ll m_{H}^{2}$ limit of the partial-wave amplitudes $a_{IJ}(s)$
\cite{DV}, which turns out to be valid for a wide range of energy, to find the
Landau ghost. The contribution of the fermion loop to these low-energy
amplitudes can be summarized as follows
\begin{eqnarray}
a_{IJ}^{f}(s)=\frac{s^2}{512\pi^{3}v^{4}}N_{c}E_{IJ}(r)\,,\quad r=\frac{M_{H}^{2}}{m_{f}^{2}}
\label{alow}
\end{eqnarray}
where $E_{IJ}(r)$ is a channel dependent algebraic factor and $N_{c}$ is
the number of color or fermion species. We shall refer loosely to
$N_{c}E_{IJ}(r)$ as the ''effective number of fermions'' in the $IJ$
channel. In the limit of the non linear
sigma model (NLSM)\cite{Lehmann}, $r\rightarrow\infty$, we have
$E_{00}(\infty)=1/3\,,\ E_{11}(\infty)=2/9\,,\ E_{20}(\infty)=2/3$ and
these numbers are all positive and small. For finite $r$ this is no
longer the case and the effective numbers of fermions, which vary
wildly with the mass ratio $r$, can be large and negative. We find that:
\begin{eqnarray}
\label{ratios}
\lim_{r\rightarrow
0}\,\frac{rE_{IJ}(r)}{E_{IJ}(\infty)}=-\frac{176}{3},\,8,\,-\frac{16}{3}
\end{eqnarray}
for $a_{00}^{f}$, $a_{11}^{f}$ and $a_{20}^{f}$ amplitudes respectively.
These limits are a good approximation already when $r$ is of
order unity or less. On the contrary the NLSM limit can only be
reached when $r$ is much larger than unity. For example the value
$E_{00}=0.2$ (compared with the NLSM value of 1/3) is reached only when $r=40$.

        The main problem with perturbation theory comes from
the large effective negative number of fermions in the $I=J=0$ channel
when $r \leq 1$.
Such a large negative number  will inevitably introduce a Landau
ghost close to origin. This is an essential point.
In Table~\ref{ghosts} we give the position of the Landau ghost as a
function of fermion and Higgs masses. The result from our calculation is very
unsatisfactory: As can be seen from this table, for most values of
$r$, equal or less than unity, the Landau ghost moves towards the
origin along the negative s-axis. Because our calculation should be
a good low energy calculation, the presence of the low energy Landau
ghost is therefore not acceptable.
        
The $I=2,\,J=0$ channel has a similar problem. For $r\leq 1$
the effective fermion number for this channel is also negative. For 
$r<1$ the Landau ghost in this channel moves near the real positive s-axis.
It gives rise to a ''resonance'' in this channel with a negative scattering
length and negative effective range which is completely unphysical.
It is not clear whether this undesirable solution is
due to the Pad\'e scheme or to some thing more fundamental which is related
to a negative effective fermion number.
        
Unlike the S-wave scattering, the P-wave channel $I=J=1$
has the effective fermion number positive for all value of $r$. In this case
there is no Landau ghost present in our solution even with fermion and Higgs
values which give difficulties in the S-wave channel.

        In summary, we find that the Landau ghosts are far from the physical
region of interest provided that the ratio $r \gg 1$. Although we found
satisfactory solutions for the $I=J=1$ channel even for $r=1$ or
less, the S-wave solutions are very unsatisfactory. One could speculate that
there could be some fundamental difficulties in a theory where  fermion
masses are larger or equal to the scalar masses or simply that we have not
treated the channel $W_{L}W_{L}\rightarrow F\bar{F}$ in a satisfactory manner.

\section{Conclusion}

        We have made a detailed study of the strongly interacting
Higgs with or without the presence of the heavy fourth generation.
Without the fourth generation, our conclusion is very much the
same as given in the previous study \cite{Ghost}: It is not possible
to generate a P-wave resonnance in $W_{L}W_{L}$ scattering with a mass
less than 3 TeV. A heavy Higgs boson with a mass larger than 3 TeV
would generate a lower mass Higgs boson corresponding to the $I=J=0$
phase shift passing through $90^{\circ}$ from below which should be considered 
as as a genuine resonance. The input Higgs boson mass corresponds to a
phase shift passing through $90^{\circ}$ from larger
values (antiresonance). Yet a consistent pertubative calculation
must be done with the input parameter
otherwise the chiral properties are spoiled. Here there is a potential
difficulty with the perturbative scheme.
It is amusing to note that only in potential scattering theory one can  prove 
that the rate of decrease of the phase shift through $90^{\circ}$ is
related through causality to the range of interaction. The
existence of an anti-resonance cannot be excluded a priori. 

        When the fourth generation is introduced, the problem becomes
much more complicated due to the large effective number of fermions. For
the ratio of Higgs mass to fermion mass of the order of unity, the
P-wave fermion number is positive and is enhanced by a factor of 8, the
$I=J=0$  is enhanced by a factor of 60, the $I=2$, $J=0$ is of the order of 2
and they are both negative. Because of the P-wave enhancement factor,
the expected mass is much lower than that obtained from the NLSM;
furthermore, because the tree $I=J=1$ amplitude decreases from the low
energy theorem  value as $(1-s/m_{H}^{2})$ the P-wave resonance has a
much narrower width. 

        In the $I=J=0$ channel, because of the enormous  negative enhancement
factor, the Landau ghost is driven to the origin, which makes the resulting
phase shift unacceptable. The same situation holds also for the $I=2$,
$J=0$ channel. Here we have an even more difficult situation: when the
fermion mass is larger or equal to the Higgs mass, a resonance is
generated in this channel corresponding to a Landau ghost of the same
mass, which is slightly off the real energy axis and on the physical
sheet. Unlike the usual result of the screening theorem of Veltman,
the presence of the heavy fermion gives rise to a strong dependence on
the Higgs mass or more precisely on the ratio of the Higgs boson to
the fourth family fermion masses. Our main conclusion is that we can
get a reasonable solution with heavy fermions of the order of less
than 0.5 Tev provided that the corresponding Higgs boson is at least
twice as heavy. We can also incorporate heavier fermion masses
provided we take heavier Higgs mass.

\vskip 1.0cm
\leftline{\bf Acknowledgement}
\vskip 0.5cm
One of us (TNT) would like to thank Ray Willey for useful
discussions. LVD and G.~Jikia would like to thank the Centre de
Physique Th\'eorique de l'Ecole Polytechnique for hospitality.

\begin{table}
\center
\small
\begin{tabular}{|c|c||c|c|c|} \hline
$m_{H}$, TeV & $m_{f}$, TeV & $I=0,\,J=0$ & $I=1,\,J=1$ & $I=2,\,J=0$\\
\hline\hline
     & 0   &        &      &      \\
     & 0.5 & 0.78   & 1.11 &      \\
     & 0.7 & $<$ 0.78 & 1.68 &      \\
 0.5 & 1   & 1.14   & 1.89 & 1.89 \\
     & 1.5 & 0.60   & 1.68 & 1.21 \\
     & 2   & 0.60   & 1.37 & 0.95 \\
\hline
     & 0   &        &      &      \\
     & 0.5 & 0.97   & 1.13 &      \\
     & 0.7 & $<$ 1.2  & 1.70 &      \\
 0.7 & 1   & 1.20   & 1.89 & 1.94 \\
     & 1.5 & 0.90   & 1.70 & 1.28 \\
     & 2   & 0.90   & 1.41 & 1.03 \\
\hline
     & 0   &        &      &      \\
     & 0.5 &        & 1.17 &      \\
     & 0.7 &        & 1.74 &      \\
 1   & 1   &        & 1.89 & 2.02 \\
     & 1.5 &        & 1.73 & 1.40 \\
     & 2   &        & 1.48 & 1.14 \\
\hline
     & 0   &        &      &      \\
     & 0.5 &        & 1.22 &      \\
     & 0.7 &        & 1.89 &      \\
 1.5 & 1   &        & 1.89 & 2.19 \\
     & 1.5 &        & 1.80 & 1.59 \\
     & 2   &        & 1.59 & 1.32 \\
\hline
     & 0   & 1.36   &      &      \\
     & 0.5 & 1.27   & 1.19 & 1.30 \\
     & 0.7 & 1.28   & 1.69 & 1.27 \\
  5  & 1   & 1.36   & 1.96 & 1.30 \\
     & 1.5 & 1.71   & 2.09 & 1.57 \\
     & 2   & 1.19   & 1.97 &      \\
\hline
     & 0   & 1.27   &      & 1.25 \\
     & 0.5 & 1.21   & 1.19 & 1.09 \\
     & 0.7 & 1.21   & 1.65 & 1.07 \\
 10  & 1   & 1.23   & 1.98 & 1.06 \\
     & 1.5 & 1.25   &      & 1.06 \\
     & 2   & 1.31   &      & 1.08 \\
\hline
\end{tabular}
\normalsize
\caption{Scale $\Lambda_{c} $ of strong interactions in the higgs sector of the
standard model as a function of the mass parameters $m_{H}$ and
$m_{f}$ ($m_{f}=0$ means no fermions). Empty slots
 mean that $\Lambda_{c}$ is far above the inelastic
threshold or near the Higgs pole.
}
\label{scales}
\end{table} 

\begin{table}
\center
\begin{tabular}{|c|c||c|c|c|} \hline
$m_{H}$, TeV & $m_{f}$, TeV & $I=0,\,J=0$, TeV$^{2}$ & $I=2,\,J=0$, TeV$^{2}$\\
\hline\hline
      & 1.5 &  (-0.339, 0.014) & ( 1.059, 0.163) \\
 1.5  & 1.  &  (-0.694, 0.055) & ( 2.188, 0.659) \\
      & 0.7 &  (-1.017, 0.113) & ( 3.095, 1.289) \\
      & 0.5 &  (-1.777, 0.321) & ( 4.752, 3.025) \\
\hline
      & 1.5 &  (-0.150, 0.003) & ( 0.440, 0.030) \\
      & 1.  &  (-0.326, 0.013) & ( 0.975, 0.139) \\
  1.  & 0.7 &  (-0.602, 0.042) & ( 1.771, 0.439) \\
      & 0.5 &  (-0.898, 0.089) & ( 2.526, 0.870) \\
      & 0.3 &  (-1.845, 0.345) & \\
\hline
      & 1.5 &  (-0.073, 0.001) & ( 0.209, 0.007) \\
      & 1.  &  (-0.162, 0.003) & ( 0.469, 0.033) \\
 0.7  & 0.7 &  (-0.316, 0.012) & ( 0.912, 0.122) \\
      & 0.5 &  (-0.553, 0.035) & ( 1.546, 0.338) \\
      & 0.3 &  (-0.929, 0.095) & \\
\hline
      & 1.5 &  (-0.037, 0.000) & ( 0.104, 0.002) \\
      & 1.  &  (-0.083, 0.001) & ( 0.235, 0.009) \\
 0.5  & 0.7 &  (-0.167, 0.003) & ( 0.471, 0.034) \\
      & 0.5 &  (-0.308, 0.011) & ( 0.858, 0.109) \\
      & 0.3 &  (-0.663, 0.050) & \\
\hline
 0.2  & 0.2 &  (-0.286, 0.010) & ( 0.740, 0.081) \\
\hline                   
\end{tabular}
\caption{Positions of the complex Landau ghosts as a function of the
  mass parameters $m_{H}$ and $m_{f}$ in the $I=0$ and $I=2$ channels
using the low-energy expansions of the partial-wave amplitudes.
}
\label{ghosts}
\end{table}                                                    

\begin{figure}
\epsfig{file=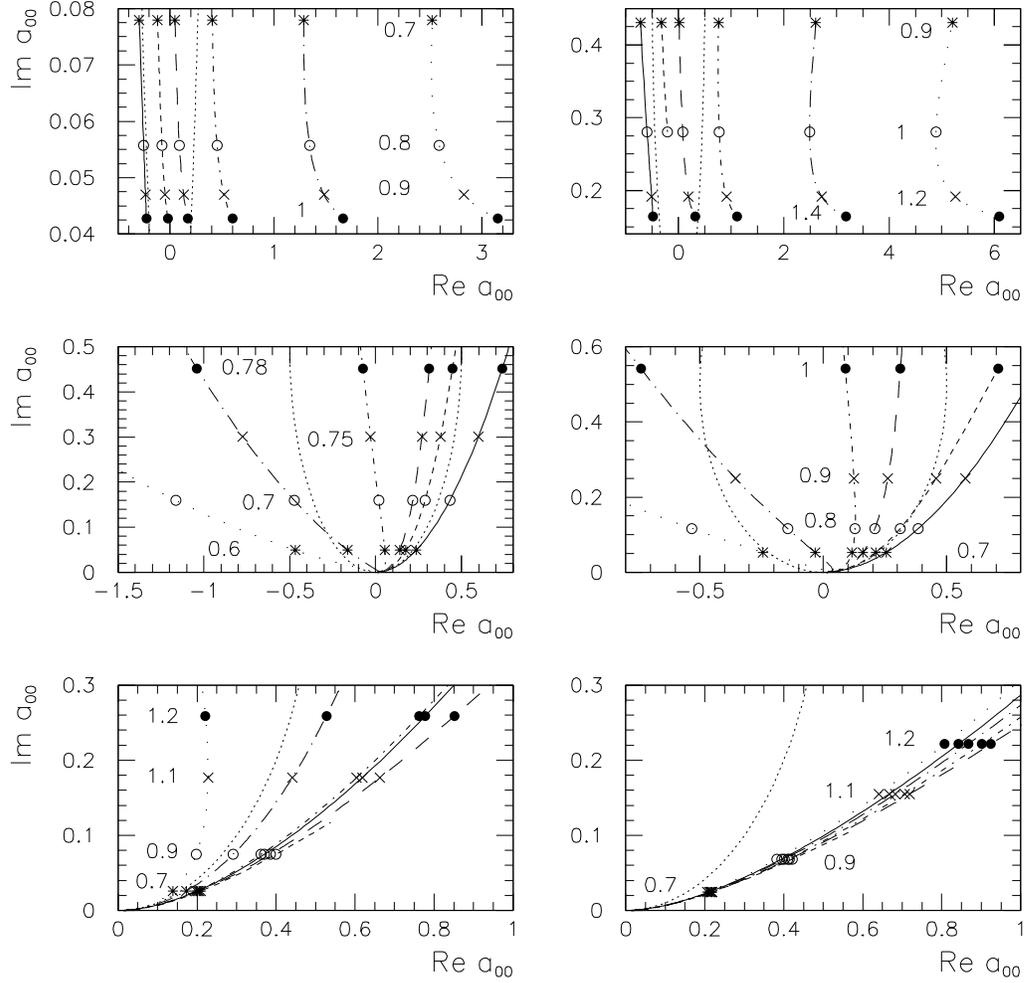,width=15cm}
\caption{Argand plots for $I=0$, $J=0$ channel for $m_H=0.5$, 0.7, 1, 1.5, 5
and 10~TeV. Solid line corresponds to the case of no fermion contribution,
short dashed line to $M_f=0.5$~TeV, long dashed line to $M_f=0.7$~TeV,
short dash-dotted line to $M_f=1$~TeV, long dash-dotted line to $M_f=1.5$~TeV
and long dotted line to $M_f=2$~TeV. Also unitarity circle is shown by short
dotted line. Markers denote energy values in TeV.}
\end{figure}

\begin{figure}
\epsfig{file=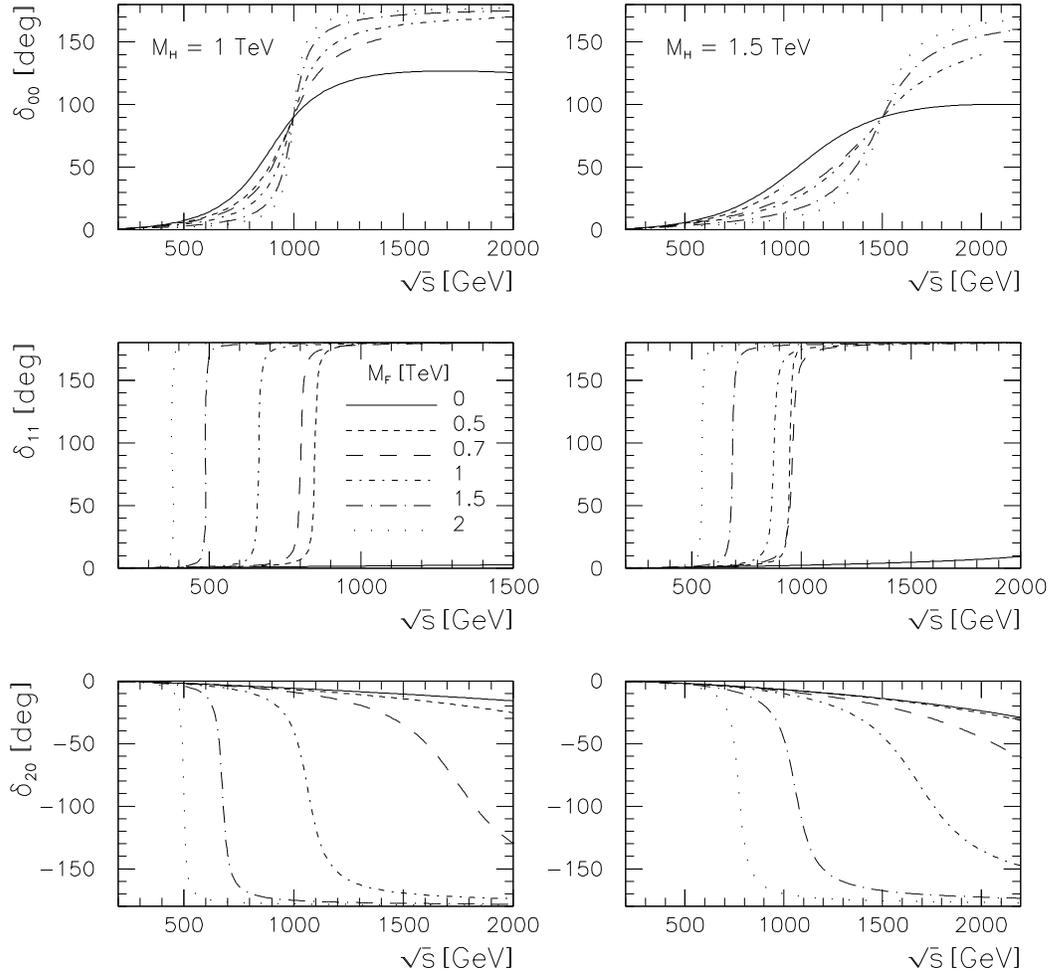,width=15cm}
\caption{Phase shifts for $m_H=1$ and 1.5~TeV as a function of energy for
various fermion masses.}
\end{figure}

\end{document}